\documentclass[useAMS,usenatbib]{mn2e}
\usepackage{graphicx}
\usepackage{times}

\def\HI {H\,{\sc i}}                       
\def\kms {~km\,s$^{-1}$}                   
\def\Lo {~L$_\odot$}                       
\def\Mo {~M$_\odot$}                       
\def\FHI {$F_{\rm HI}$}                    
\def\MHI {$M_{\rm HI}$}                    
\def\NHI {$N_{\rm HI}$}                    
\def\ccm {cm$^{-3}$}                       
\def\scm {cm$^{-2}$}                       
\def\MOLH {H$_2$}                          
\def\WAT {H$_2$O}                          
\def\rr {\rightarrow}                      
\def\AB {$A_{\rm B}$}                      
\def\LB {$L_{\rm B}$}                      
\def\lapp{\ifmmode\stackrel{<}{_{\sim}}\else$\stackrel{<}{_{\sim}}$\fi}
\def\gapp{\ifmmode\stackrel{>}{_{\sim}}\else$\stackrel{>}{_{\sim}}$\fi}

\title[Large-Scale Gas in the Circinus Galaxy]
      {The Large-Scale Atomic and Molecular Gas in the Circinus Galaxy}
\author[S. J. Curran,  B. S. Koribalski \& I. Bains]
       {S. J. Curran$^{1,2}$\thanks{E-mail: sjc@phys.unsw.edu.au},
	B. S. Koribalski$^{3}$ and I. Bains$^{4}$\\
$^{1}$School of Physics, University of New South Wales, 
      Sydney NSW 2052, Australia\\
$^{2}$Onsala Space Observatory, Chalmers University of Technology, 
      S-439 92 Onsala, Sweden\\
$^{3}$Australia Telescope National Facility, CSIRO, 
      PO Box 76, Epping NSW 1710, Australia\\
$^{4}$Centre for Astrophysics and Supercomputing, Swinburne University
of Technology, PO Box 218, Hawthorn, VIC 3122, Australia
}

\begin{document}

\date{Accepted ---. Received ---; in original form ---}

\pagerange{\pageref{firstpage}--\pageref{lastpage}} \pubyear{2008}

\maketitle

\label{firstpage}

\begin{abstract}
We have used the Australia Telescope Compact Array (ATCA) and the
Swedish-ESO Sub-millimetre Telescope (SEST) to map the large-scale
atomic and molecular gas in the nearby (4~Mpc) Circinus galaxy. The
ATCA \HI\ mosaic of Circinus exhibits the warps in position angle and
inclination revealed in the single-pointing image of \citet{jkeh98},
both of which appear to settle beyond the inner 30 kpc which was
previously imaged. The molecular gas has been mapped in both the CO
$J=1\rr0$ and $J=2\rr1$ transitions down to a column density of
$N_{{\rm H}_{2}}\ga10^{21}$ \scm ~($3\sigma$), where we derive a total
molecular gas mass of $M_{\rm H_2} \approx2 \times 10^9$\Mo. Within a
radius of 3 kpc, i.e. where CO was clearly detected, the molecular
fraction climbs steeply from $\approx0.7$ to unity (where $N_{{\rm
    H}_{2}}=4\times10^{22}$ \scm, cf. $N_{\rm HI}=10^{21}$ \scm) with
proximity to the nucleus. Our \HI\ mosaic gives an atomic gas mass of
$M_{\rm HI} \approx6 \times 10^9$\Mo, which is 70\% of the fully
mapped single dish value. Combining the atomic and molecular gas
masses gives a total gas mass of $M_{\rm gas}\equiv M_{\rm HI} +
M_{\rm H_{2}}\approx1\times10^{10}$\Mo, cf. the total dynamical mass
of $\approx3\times10^{11}$\Mo\ within the inner 50 kpc of our mosaiced
image.  The total neutral gas mass to dynamical mass ratio is
therefore 3\%, consistent with the SAS3 classification of
Circinus. The high (molecular) gas mass fraction of $M_{\rm
  H_2}/M_{\rm dyn}\approx50$\% found previously \citep{cjrb98}, only
occurs close to the central $\approx0.5$ kpc and falls to $\lapp10$\%
within and outwith this region, allaying previous concerns regarding
the validity of applying the Galactic $N_{\rm H_{2}}/I_{\rm CO}$
conversion ratio to Circinus. The rotation curve, as traced by both
the \HI\ and CO, exhibits a steep dip at $\approx1$ kpc, the edge of
the atomic/molecular ring, within which the star-burst is
occurring. We find the atomic and molecular gases to trace different
kinematical features and believe that the fastest part ($\gapp130$
\kms) of the sub-kpc ring consists overwhelmingly of molecular
gas. Beyond the inner kpc, the velocity climbs to settle into a solid
body rotation of $\approx150$ \kms\ at $\gapp10$ kpc. Most of the
starlight emanates from within this radius and so much of the
dynamical mass, which remains climbing to the limit of our data
($\ga50$ kpc), must be due to the dark matter halo.
\end{abstract}

\begin{keywords}
   galaxies: active --- galaxies: individual: Circinus --- galaxies:
   kinematics and dynamics --- galaxies: ISM --- galaxies: Seyfert ---
   radio lines: galaxies.
\end{keywords}

\section{Introduction} 
\label{sec:intro}

Circinus is a nearby, highly inclined spiral galaxy located behind the
Galactic Plane (at $l,b$ = 311\fdg3, --3\fdg8), where the high stellar
density and dust extinction (\AB\ = 6.3 mag, \citealt{sfd98}) make it
difficult to study its optical properties and environment. At a
distance of only $4.2 \pm 0.8$ Mpc \citep{fkl+77}, Circinus represents
one of the closest examples of a star-burst galaxy with a Seyfert\,2
nucleus \citep{mo90,osmm94}. 
Low-resolution \HI\ observations of Circinus with the 64-m Parkes
radio telescope revealed a vast galaxy with \HI\ 21-cm emission
extending to a half-width of 32\arcmin\ \citep{fkl+77}. Its enormous
extent has also been noted in both the shallow and deep \HI\ surveys
of the Zone of Avoidance (HIZSS, \citealt{hse+00} and HIZOA;
\citealt{jsk+00}, respectively) and, most recently, in the \HI\
Parkes All-Sky Survey (HIPASS, \citealt{ksk+04}). 

High resolution, single-pointing ATCA \HI\ observations of Circinus
\citep{jkeh98} reveal a complex velocity field and gas
distribution out to a diameter of $\sim80$\arcmin\ ($\sim100$ kpc),
with strong kinematical warping (also evident between the large-scale
position angles and inclinations, Table \ref{info}).
\begin{table} 
\caption{General properties of the Circinus galaxy.\label{info}}
\begin{center}
\begin{tabular}{lllc} 
\hline 
central position    & $\alpha_{\rm J2000}$ &  
                     $14^{\rm h}\,13^{\rm m}\,09\fs95$ ($\pm$0\fs02) & 1 \\
                    & $\delta_{\rm J2000}$ & 
                     --65\degr\,20\arcmin\,21\farcs2 ($\pm$0\farcs1) &     \\
type                &                & SAS3                          & 2 \\
adopted distance    & $D$            & 4.2 Mpc                       & 3 \\
optical extent$^*$  & $D_{24.5}$     & $11\farcm9 \times 4\farcm6$   & 3 \\
                    & $D_{26.6}$     & $17\farcm2 \pm 1\farcm7$    & 3 \\ 
position angle      & $PA_{\rm opt}$ &  220\degr                     & 2 \\
                    & $PA_{\rm CO}$  & $214\degr \pm 4\degr$         & 4 \\
                    & $PA_{\rm HI}$  & $210\degr \pm 5\degr$         & 3 \\
inclination         & $i_{\rm opt}$  &  64\degr                      & 2 \\
                    & $i_{\rm CO}$   & $78\degr \pm 1\degr$  ($<30$\arcsec) & 4 \\
                    & $i_{\rm HI}$   & $65\degr \pm 2\degr$          & 3 \\
systemic velocity   & $v_{\rm HI}$   & $439 \pm 2$\kms\              & 3 \\
\HI\ flux density   & \FHI\          & $1910 \pm 130$ Jy\kms         & 3 \\
\HI\ mass           & \MHI\          & $7.9\pm0.5\times10^{9}$\Mo    & 3 \\
dynamical mass      & $M_{\rm dyn}$  & $1.3\pm0.2\times10^{11}$\Mo   & 3 \\
$B$-band luminosity & \LB\           & $7  \pm2  \times10^{9}$\Lo    & 3 \\
$B$-band extinction & \AB\           & 6.3 mag                       & 5 \\
\hline
\end{tabular}
\end{center}
Notes:
   Freeman et al. (1977) derived extinction-corrected optical diameters
   for the Circinus galaxy at the 24.5th blue magnitude, $D_{24.5}$, and, 
   through extrapolation, the 26.6th blue magnitude, $D_{26.6}$ (the Holmberg diameter).\\ 
 References:
   (1) \protect\citet{gbe+03}, 
   (2) \protect\citet{ddc+91}, 
   (3) \protect\citet{fkl+77}, 
   (4) \protect\citet{cjrb98},
   (5) \protect\citet{sfd98}.
\end{table}
The atomic gas is distributed in a $\sim$10~kpc radius ring enclosing
a bar, in turn terminating in a 1 kpc radius ring \citep{jkeh98}. This
inner \HI\ ring may be an outer component of the 600 pc CO ring
\citep{cjrb98}, where the star formation is dominant
\citep{mmoo94}. This itself may feed the nucleus via a nuclear bar
nestled within the molecular ring \citep{maa+00}, thus providing a
means of transporting gas from the outer galaxy to the central engine
(e.g. \citealt{sfb89}).

The notion of a continuous structure from the kpc to sub-pc scale is
supported by the VLBI observations of the \WAT\ maser emission
\citep{gbe+03} and VLTI near infrared observations \citep{tmj+07},
which both show that the sub-pc accretion disk/obscuring torus shares a
similar close to edge-on orientation to the larger-scale molecular gas
\citep{cjrb98}.  The inner edge of the warped maser disk shares a
similar position angle to the large-scale gas disk ($PA = 209\degr \pm
3\degr$), although this increases to $236\degr \pm 6\degr$ at the
outer edge \citep{gbe+03}. Furthermore, a second population of masers
are found to trace a wide angle outflow which coincides with the
$>$100 pc-scale ionisation cone (Marconi et al. 1994, \citealt{vb97}) and
molecular outflow (\citealt{crjb98}), as well as sharing the same
position angle as the radio lobes \citep{hwr+90,ehj+95}. The lobes,
cone and outflow are all directed along the minor axis of the galaxy,
coincident with the rotation axis of the molecular ring
\citep{crjb98,cur01}.

Previously, \citet{jkeh98} obtained very sensitive, single-pointing
\HI\ data of the Circinus galaxy with the Australia Telescope Compact
Array (ATCA). Since the ATCA primary beam is $\sim$33\arcmin, these
data were not sufficiently sensitive to detect the outer edges of the
large gaseous disk of Circinus. Here we present and analyse an \HI\
mosaic of Circinus, obtained with the 375-m configuration of the ATCA,
covering the whole galaxy. This is complemented by the most extensive
CO~$1\rr0$ and $2\rr1$ maps of the galaxy to date.

\section{Observations and data reduction}\label{obs} 

\subsection{\HI\ observations}\label{hiobs} 

The \HI\ observations were performed with the ATCA\footnote{The
  Australia Telescope is funded by the Commonwealth of Australia
  for operations as a National Facility managed by CSIRO.}  in its
  375-m configuration, on June 18--22, 1999. In order to avoid solar
  interference on the short baselines of the array, which are needed
  to detect the very extended \HI\ emission, we only observed at
  night. We mosaiced Circinus using a grid of $5 \times 3$ points
  along the major axis with pointing centres separated by 16\farcm5
  (i.e. half the primary beam-width) in order to fully Nyquist sample
  the field. The band was centered at 1418 MHz with a bandwidth of 8
  MHz, divided over 512 channels, giving a channel spacing of 3.3\kms,
  i.e. a velocity resolution of 4\kms. We observed for $5 \times 12$
  hours in total, achieving an integration time of 44 hours on
  Circinus, i.e. $\approx$3 hours per pointing. We used PKS B1934--63
  for flux and bandpass calibration and PKS B1329--665 for phase
  calibration. Data reduction and analysis were performed with the
  {\sc miriad} software package.

After identifying the velocity range of \HI\ emission in Circinus and
noting the foreground emission from the Galaxy and high velocity
clouds, we subtracted an average of the line-free channels from the
{\em uv}-data.  The \HI\ data were then Fourier-transformed using
`natural' weighting. We used ten baselines between 31\,m and 459\,m,
excluding the five longer baselines to the distant antenna six, which
do not contribute to the \HI\ emission. The task {\sf mossdi} was used
to clean the mosaiced \HI\ cube which was then restored with a beam
size of $124\arcsec \times 107\arcsec$. The r.m.s. noise per each 4\kms\
channel is $\approx2$~mJy\,beam$^{-1}$, close to the theoretical
value.

\subsection{CO observations}\label{coobs} 

The CO~$1\rr0$ and $2\rr1$ observations of the inner $\sim8\arcmin \times
2\arcmin$ region of the Circinus galaxy were performed with the
SEST\footnote{Operated until 2003 by the European Southern Observatory
(ESO) and the Swedish National Facility for Radio Astronomy, Onsala
Space Observatory, Chalmers University of Technology.} at La Silla,
Chile. In total, a grid of $7 \times 27$ points separated by
20\arcsec\ was observed along the major axis and in $7 \times 13$ points
along the minor axis. The integration time was 15 minutes per grid
point (in 3 minute scans), giving a typical r.m.s. of $T_A^*\lapp30$ mK
per 0.7 MHz channel. On June 7--9, 1999, we observed the central three
strips, at position offsets of --20\arcsec, 0 \& +20\arcsec\ along
both the major and minor axes and on April 22, 2000, we observed the
remaining positions.  The CO~$1\rr0$ and $2\rr1$ transitions were
mapped simultaneously with the 115 \& 230 GHz (IRAM) receivers, tuned
to single-sideband mode. Typical system temperatures, on the
$T_A^*$-scale, were 300 -- 400 K at 115 GHz and 300 K at 230 GHz. The
back-ends were acousto-optical spectrometers with 1440 channels and a
channel width of 1.8\kms. We used dual-beam switching with a
throw of about 12\arcmin\ in azimuth, and pointing errors, using the
SiO maser W Hydra, were typically 3\arcsec\ r.m.s. on each axis. The
half power beam-widths (HPBWs) are 45\arcsec\ for CO~$1\rr0$ and
22\arcsec\ for CO~$2\rr1$. The intensity was calibrated using the
chopper-wheel method. For all of the observing runs the weather was
excellent, and only the removal of linear baselines, using the {\sc
xs} package, was required.

\section{Results} 

\subsection{Gas distributions}

\subsubsection{The \HI\ maps}\label{himaps} 
\begin{figure*} 
\vspace{11.0cm}
\includegraphics{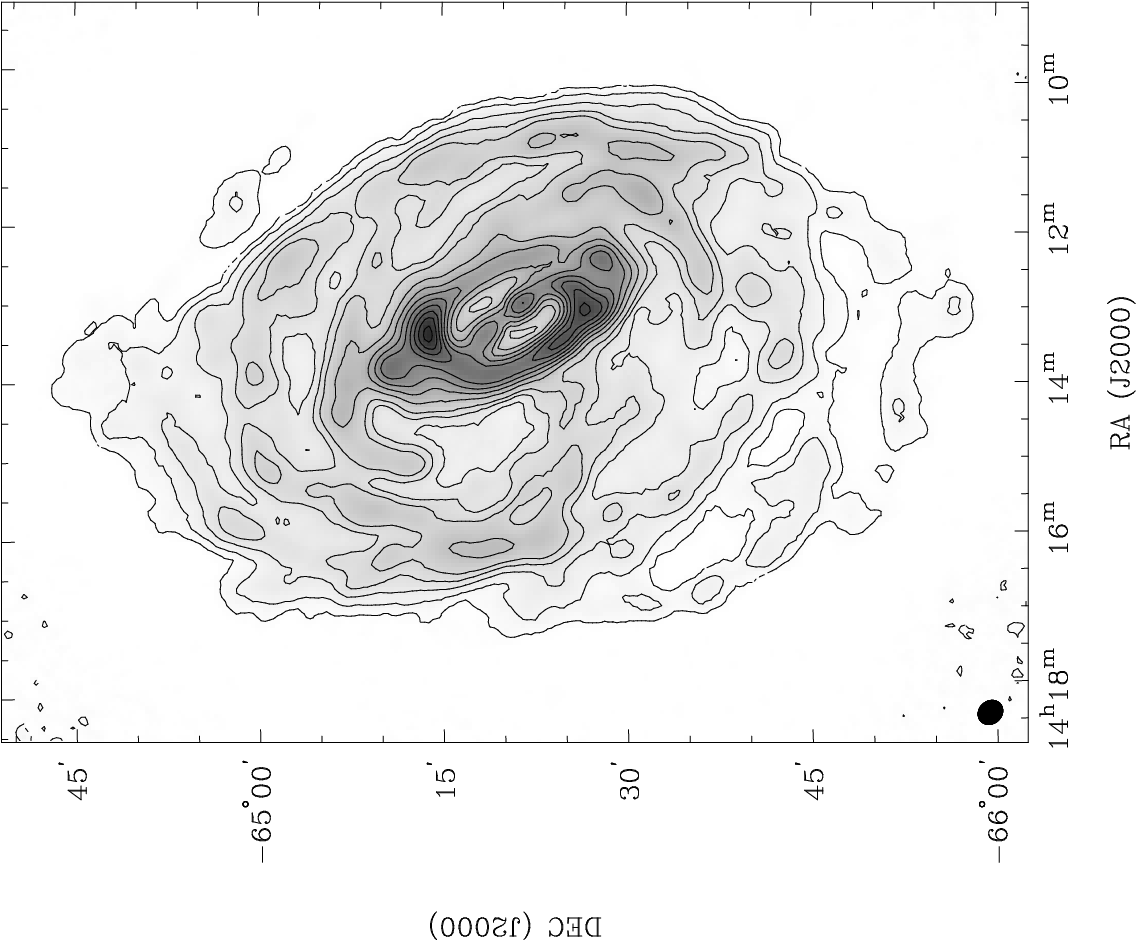}
\includegraphics{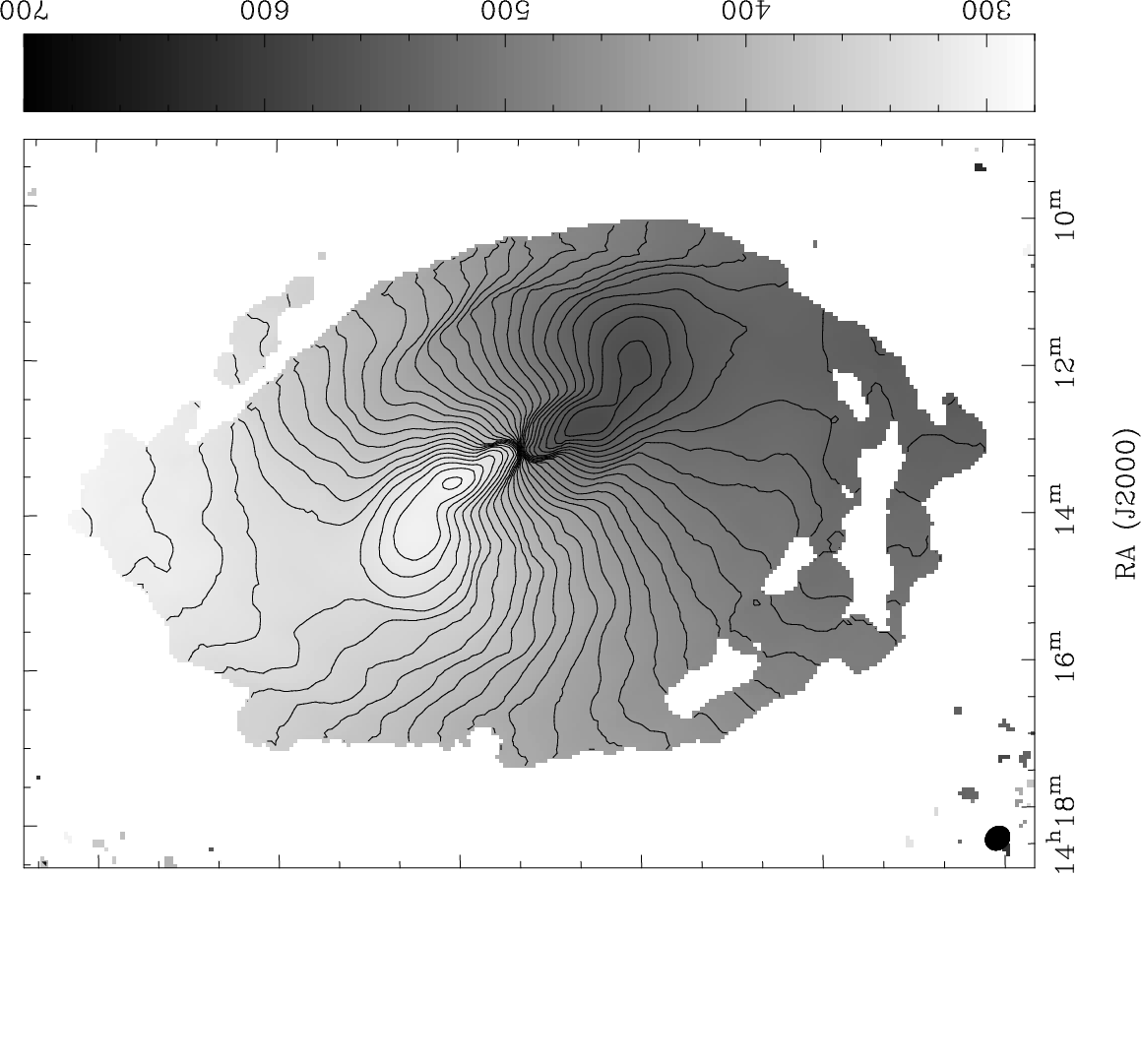}
\caption{High-resolution \HI\ moment maps of the Circinus galaxy. 
  Left: The \HI\ distribution; the contour levels are 0.4, 1, 2, 3, 4, 
  6, 8, 10, 12, 18 and 20 Jy\kms\,beam$^{-1}$. 
   Right: The mean \HI\ velocity field; the contour levels range from 
  300\kms\ (NE) to 570\kms\ (SW) in steps of 10\kms. The synthesised beam 
  ($124\arcsec \times 107\arcsec$) is shown in the bottom left corner.}
\label{large}
\vspace{9.0cm}
\includegraphics{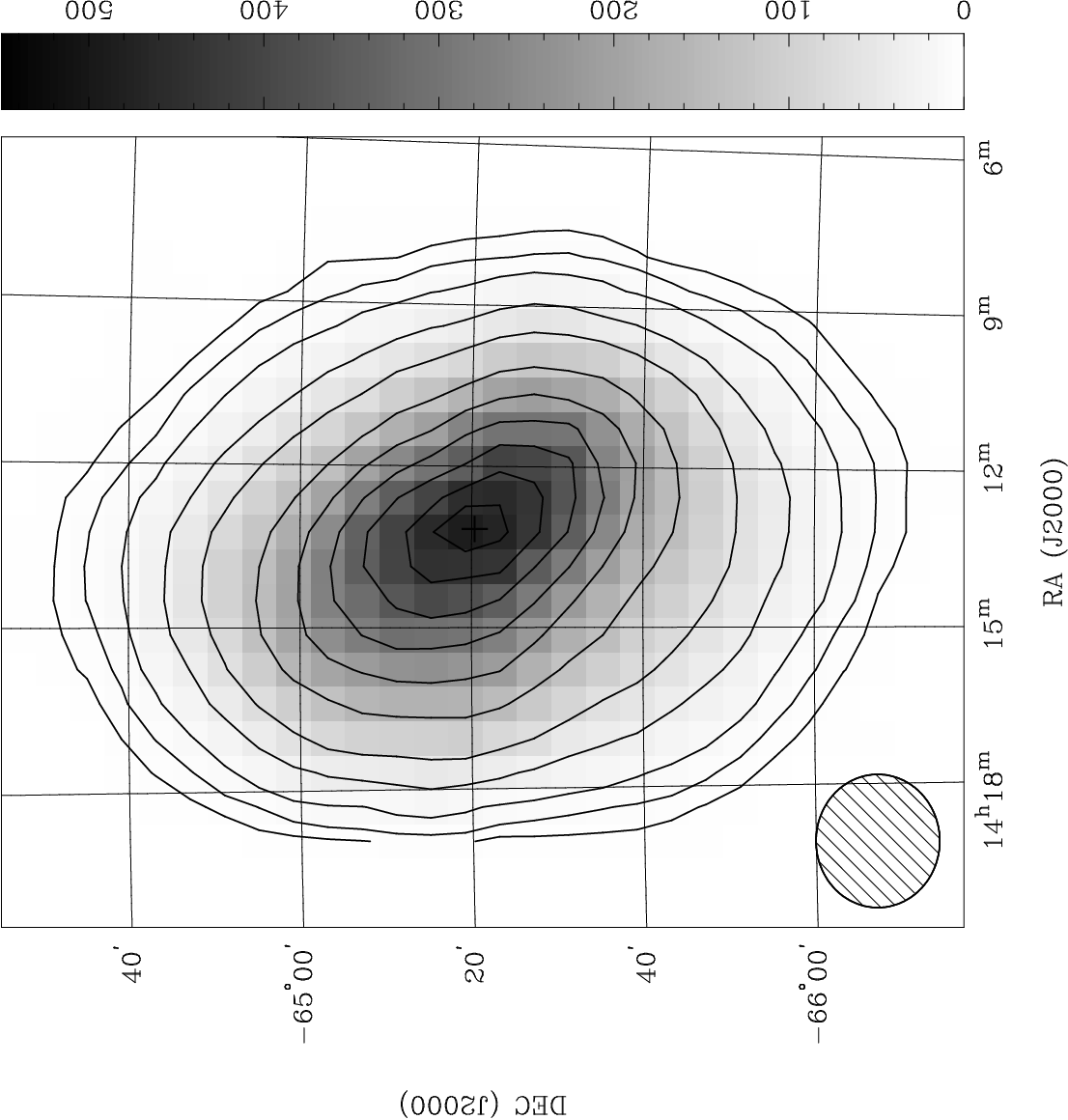}
\includegraphics{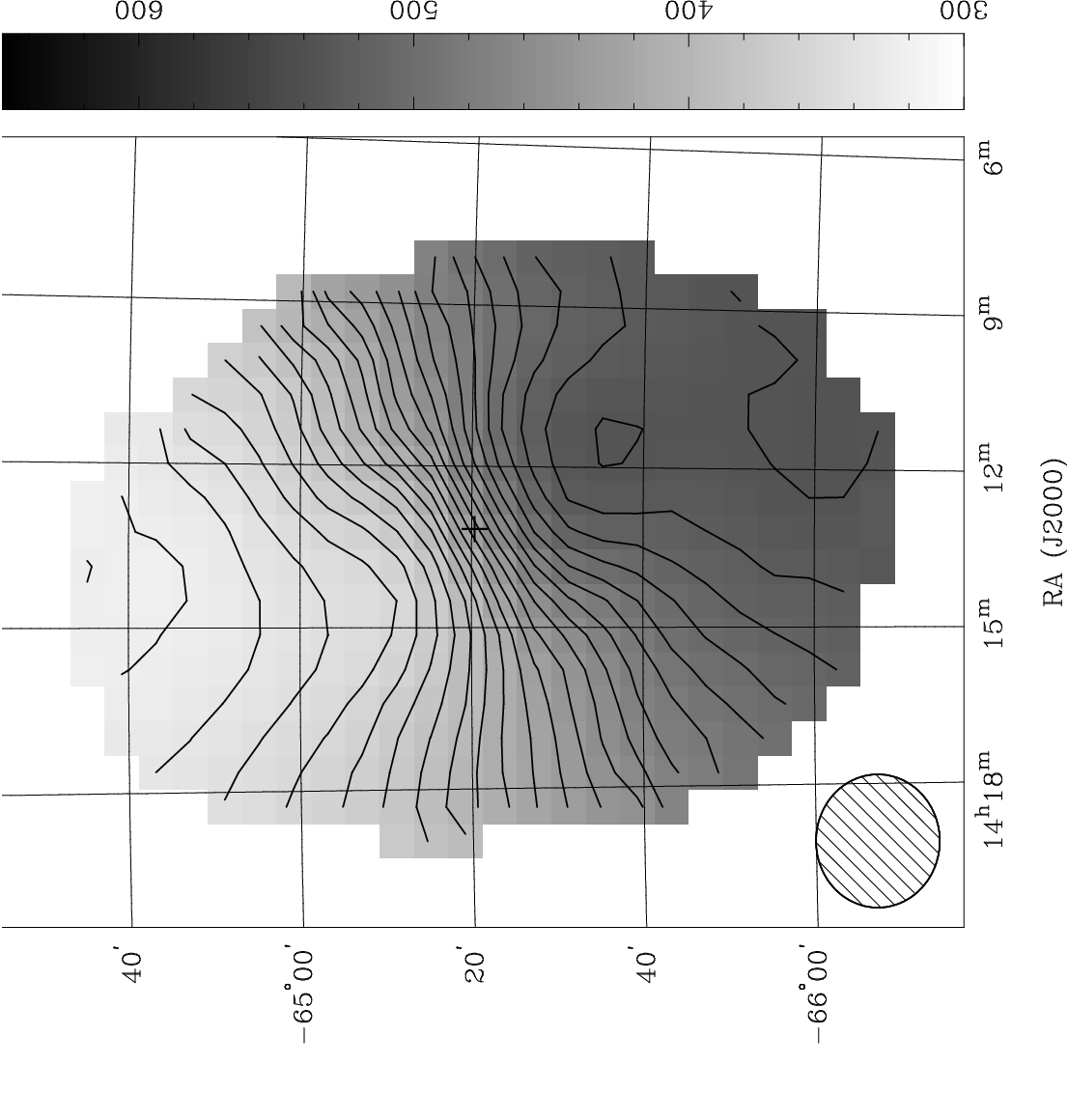}
\caption{Low-resolution \HI\ moment maps of the Circinus galaxy as
obtained from the deep Parkes \HI\ survey of the Zone of Avoidance
(\citealt{jsk+00}; Henning et al., in prep.).  Left: The \HI\
distribution; the contour levels are (0.5, 1, 2, 5, 10, 20, 30, 40,
50, 60 and 66) $\times$ 7 Jy\kms\,beam$^{-1}$, where 7
Jy\kms\,beam$^{-1}$ correspond to an \HI\ column density of $10^{19}$
cm$^{-2}$. The contour levels were chosen to match those by Freeman et
al.  (1977; their figure.~6) above their detection limit of \NHI\ = $5
\times 10^{19}$ cm$^{-2}$; our detection limit is a factor $\sim$10
lower.  Right: The mean \HI\ velocity field (masked at \NHI\ = $5
\times 10^{18}$ cm$^{-2}$); the contour levels range from 324\kms\
(NE) to 534\kms\ (SW) in steps of 10\kms. The cross marks the
center of the Circinus galaxy as given in Table~\ref{info}.  The
gridded beam of 15\farcm5 is shown in the bottom left corner.}
\label{zoa}
\end{figure*}

Fig.~\ref{large} shows the mosaiced, high-resolution \HI\ distribution
and mean \HI\ velocity field of the Circinus galaxy as obtained with
the ATCA. The equivalent low-resolution \HI\ images obtained from the
HIZOA survey are shown for comparison (Fig.~\ref{zoa}). Our maps
clearly show the enormous extent of Circinus, while exhibiting the
asymmetric outer envelope, radial position angle change and the inner
$\sim$30\arcmin\ of the single-pointing \HI\ image (see
\citealt{jkeh98}). 2MASS $JHK$-band images and photometry
\citep{jcc+03} show an infrared extent of $\approx16$\arcmin\ for
Circinus, similar to the optical Holmberg radius as extrapolated by
Freeman et al. (1977) [see Table \ref{info}]. The \HI\ diameter is at
least a factor of five larger than the largest estimated optical and
infrared diameters of Circinus.

As noted in previous maps \citep{fkl+77,jkeh98}, the ellipticity and
position angle changes with radius are indicative of a gentle warp. We
also note some asymmetries in the gas distribution and velocity field
with the extended \HI\ emission towards the SWW\footnote{This is more
apparent in the colour version of moment zero map, available from
http://www.phys.unsw.edu.au/$\sim$sjc/circinus/ \label{foot}}
being prominent. This pronounced emission in the more tenuous gas is
in the same direction as the extension in the CO emission (discussed
in Sect. \ref{warps}).


\subsubsection{The CO maps}\label{comaps} 

\begin{figure} 
\vspace{13.7cm}
\includegraphics{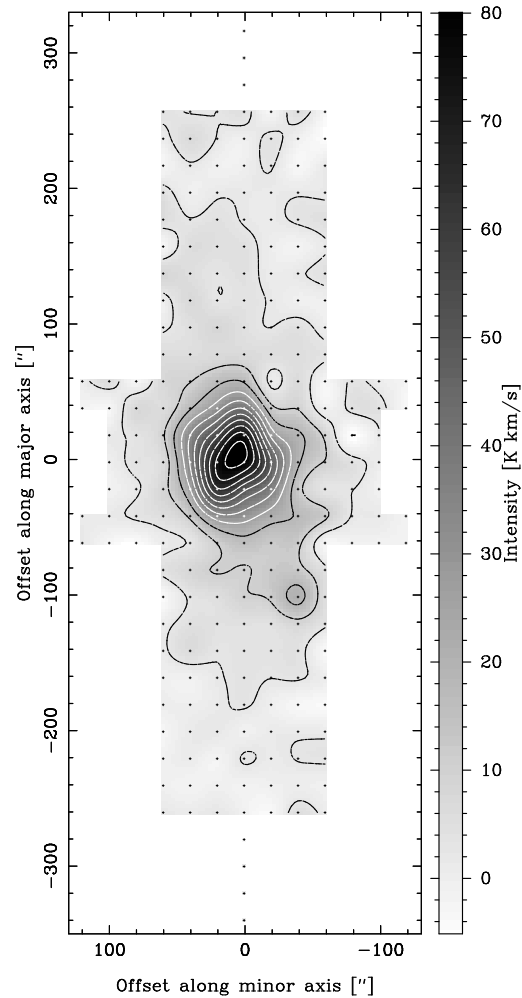}
\caption{The CO $1\rr0$ map of  Circinus. The contour levels
are up to 80 K\kms in 10\% steps. The SEST antenna gain is 27
Jy\,K$^{-1}$ at 115 GHz. In this and Fig.~\ref{co2} the position angle
is 214\degr, and the dots mark the observed positions.}
\label{co1}
\end{figure}

\begin{figure} 
\vspace{13.7cm}
\includegraphics{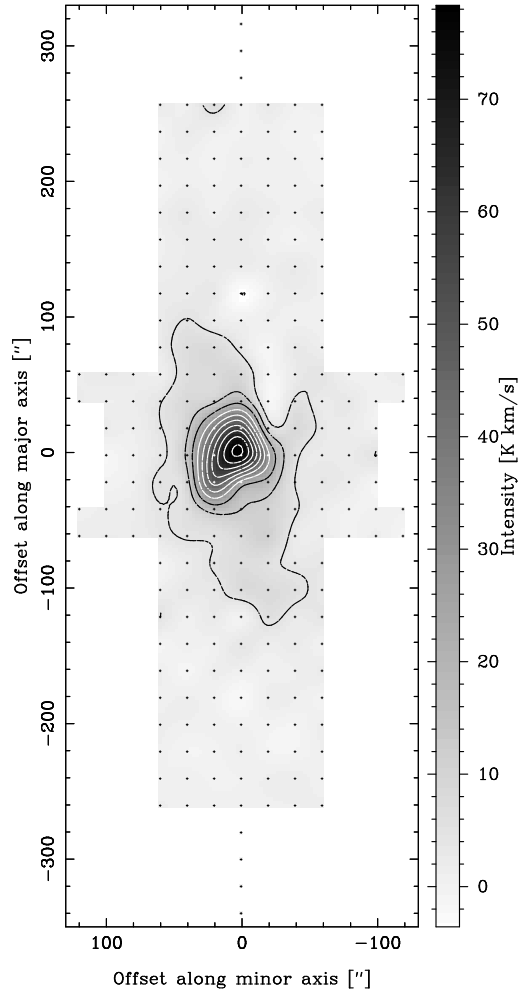}
\caption{The CO $2\rr1$ map of Circinus. The contour levels
are up to 78 K\kms ~in 10\% steps. The SEST antenna gain is 40
Jy\,K$^{-1}$ at 230 GHz.}
\label{co2}
\end{figure}

In Figs. \ref{co1} and \ref{co2} we show CO contour maps of the fully
mapped region,
where we clearly detect CO emission out to +140\arcsec\ (2.9 kpc) along the
NE ($I_{\rm A} = 4.5 \pm 1.2$ K\kms) and --160\arcsec\ (3.3 kpc) along
the SW ($I_{\rm A} = 3.8 \pm 1.1$ K\kms).  Note that, like
\citet{ekhj98}, we detect very faint emission further out along the
axes, but since these are detected at $\la2\sigma$ (e.g. $I_{\rm A} =
2.1 \pm 1.0$ K\kms\ at +180\arcsec\ along the major axis), they are
not considered further.

Along the minor axis, CO is clearly detected to 80\arcsec\ along both
the NW and SE ($I_{\rm A} = 4.8 \pm 1.3$ K\kms\ and $I_{\rm A} = 4.2
\pm 1.2$ K\kms, respectively). This deprojects to values of 3.3--6.7
kpc for $i$ = 65\degr\ to 78\degr\ (Table \ref{info}). The lower end
of this range agrees well with the value obtained from the major axis
giving $i = 60\degr \pm 6\degr$. Note, however, that within the
central 5 kpc, the \HI\ velocity field displays an elongated structure
and non-circular motions \citep{jkeh98}. This would have the effect of
lowering our estimates of the inclination of the ``disk'', although
the molecular ring+outflow model supports the high inclinations within
the central kpc, with the velocity field within this region not
exhibiting the same strong twists as on the larger scale\footnote{Due
  to the low resolution, our first moment CO maps are fairly
  featureless and the reader is referred to the over-sampled, higher
  sensitivity CO $2\rr1$ map of \citet{crjb98}.}.

CO $2\rr1$ is detected at similar main-beam temperatures as the
$1\rr0$ transition at $\approx$3.2 kpc and $\approx$1.1 kpc along the
major and minor axis, respectively, giving $i = 65\degr \pm
8\degr$. These values suggest that the large-scale CO shares a similar
inclination to the \HI\ and optical disk (Table \ref{info}), and
considering the value of $i$ = 78\degr\ found by \citet{crjb98},
indicates that the CO distribution, like the \HI\ \citep{jkeh98}, is
warped.
 
\subsection{Gas masses}
\label{gas}
\subsubsection{Atomic gas mass}
\label{agm}

\begin{figure} 
\vspace{6.4cm} 
\includegraphics{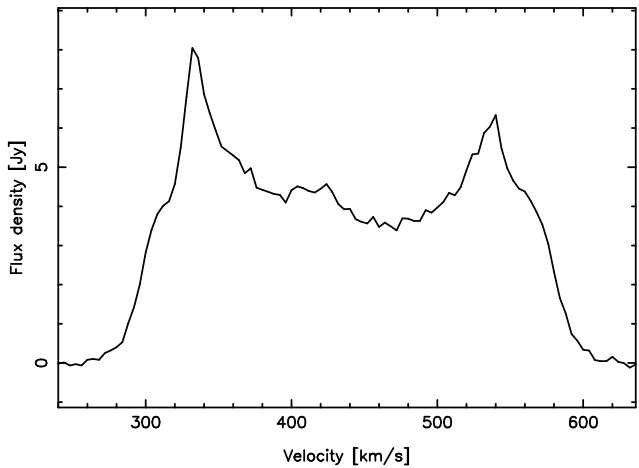}
\caption{Integrated \HI\ spectrum of the Circinus galaxy. The velocity
resolution is 4 \kms ~ and the integrated flux density is $F_{\rm HI}=
1330\pm50$ Jy \kms. All velocities quoted throughout this paper
are heliocentric.}
\label{hitotal}
\end{figure}

In Fig.~\ref{hitotal} we show the global integrated \HI\ spectrum of Circinus
obtained from our mosaic.
The integrated \HI\ flux density of $F_{\rm HI}\approx1300$ Jy
compares with $1000$ Jy \kms~from the primary beam limited image
\citep{jkeh98} and \FHI\ =1870 and 1960 Jy\kms ~from the single dish
HIZSS and HIZOA maps \citep{hse+00,jsk+00}, respectively. Our value
gives a mass of $M_{\rm HI}=5.5\pm0.2 \times10^9$~\Mo, using $M_{\rm
  HI}=2.36\times10^5\, D^2\, F_{\rm HI}$~\Mo ~\citep{wil52,rob62} and
$D=4.2$ Mpc.

Like the previous results, the spectrum shows a prominent double-horn
shape, with more emission from the fast moving (at $v\approx330$ \kms\ in
Fig. \ref{hitotal}) gas in the approaching (NE) segment. The emission peaks
occur at $332$ and $540$ \kms, giving $\pm104$ \kms\ about a centroid
of $436\pm4$ \kms. This agrees well with the systemic velocity of
$439\pm2$ \kms\ derived by \citealt{fkl+77} [see also
  Sect. \ref{sect:trm}] and indicates that the asymmetry in the
profile does not dominate the rotational dynamics of the gas.

\subsubsection{Molecular gas mass}
\label{mgm}
\begin{figure} 
\vspace{6.3cm}
\includegraphics{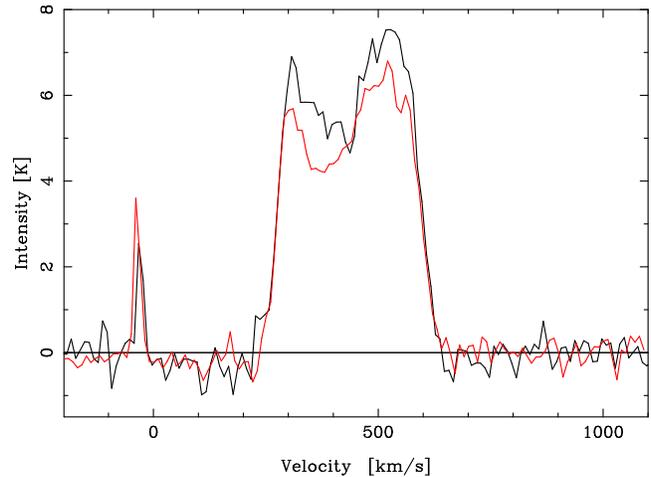}
\caption{Integrated CO $1\rr0$ (black) and $2\rr1$ (coloured) spectra
   of Circinus. The velocity resolution is 10\kms\ and the antenna
   temperatures have been corrected by main beam efficiencies of
   $\eta_{\rm mb} = 0.7$ and $0.5$, respectively.
The Galactic
   line close to 0\kms\ occurs in the western part of the map close to
   known Galactic emission \protect\citep{duc+87}, as previously noted
   by \protect\citet{ekhj98}.}
\label{cototal}
\end{figure} 
Fig.~\ref{cototal} shows the integrated CO $1\rr0$ and $2\rr1$ spectra
summed over the full map. For both transitions we see the excess of
low velocity gas in the receding peak, as was evident in the molecular
ring model of \citet{cjrb98}. This contravenes the \HI\ spectrum
(Fig.~\ref{hitotal}), indicating a significantly larger molecular
fraction in the SW portion of the ring. For CO $1\rr0$ the average
velocity integrated line intensity is $I_{\rm A} = 6.28 \pm 0.07$
K\kms. Converting this to main beam intensity, via $\eta_{\rm mb} =
0.7 \pm 10\%$ at 115 GHz, gives an average of $I_{\rm mb} = 9 \pm 1$
K\kms\ over the whole map (Fig.~\ref{co1}) resulting in a CO $1\rr0$ luminosity
of $L_{{\rm CO}~1\rr0} = 400 \pm 40$ K\kms\,kpc$^2$, where we have
added an additional half beam-width ($\Theta_{\rm mb}$) to each axis
to ensure we account for all of the mapped
region\footnote{\label{foot} For small angles the fractional
  uncertainty in this conversion is approximately equal to that of the
  distance estimate, which is $\frac{\Delta
    D}{D}=\frac{0.8}{4.2}=0.19$ (Sect. \ref{sec:intro}). This results
  in fractional errors of $\frac{\Delta M}{M} \approx 0.4$ for the gas
  mass estimates and $\approx 0.2$ for the dynamical mass estimates,
  although these uncertainties will cancel somewhat when comparing
  these masses (i.e. if one is overestimated so is the other).}.

Applying the Galactic $N_{\rm H_{2}}/I_{\rm CO}$ conversion ratio of
$2.3 \pm 0.3 \times 10^{20}$ \scm/[K \kms] \citep{sbd+88} and a 36\% mass
correction for helium (so that 1\Mo\ = $4.38 \times 10^{56}$ \MOLH\
molecules), gives a total molecular mass of $M_{\rm H_{2}} = 2.0 \pm
0.5 \times 10^9$\Mo. This is the same as the value found by
\citet{ekhj98}\footnote{They actually quote $M_{{\rm H}_{2}}=1.1\pm0.2
\times 10^9$\Mo, giving $M_{\rm H_{2}}/M_{\rm dyn}=0.05\pm0.01$, using
a $N_{{\rm H}_{2}}/I_{{\rm CO}}$ conversion ratio of $1\times10^{20}$
\scm/[K \kms]. \label{elmfoot}} and similar to that
obtained by \citet{cjrb98} over the central region only. This therefore
suggests that the previous $\pm60\arcsec \times \pm60\arcsec$ maps
contained most of the CO emission.

Since this mass is comparable with the dynamical mass, $M_{\rm dyn} =
3.3 \pm 0.3 \times 10^9$\Mo ~within 560 pc, \citet{cjrb98} raised the
possibility that the Galactic conversion ratio was inapplicable in
this case, as had been previously noted for other galaxies
(e.g. \citealt{mb88,mal90,srr94,dah95,gwt+95}). In order to check that
this apparently high molecular mass was not the result of integrating
the luminosity over $r > 560$ pc, we averaged the intensities over the
central 25 positions of Fig.~\ref{co1}, i.e. to $\pm40\arcsec\ \times
\pm40\arcsec$, which gave $I_{\rm mb} = 48 \pm 5$ K\kms.  Deconvolving
$\Theta_{\rm mb}$ from this $\pm40\arcsec$ gives a radius of 680 pc
in which a molecular mass of $M_{\rm H_{2}} = 1.0 \pm 0.2 \times
10^9$\Mo\ is found, thus either suggesting a gas mass fraction of
$\approx30$\%, within this radius, or a conversion ratio of $N_{\rm
  H_{2}}/I_{\rm CO}\la1\times10^{20}$ \scm/[K \kms].

Note that \citet{ekhj98} previously obtained a much lower molecular
gas mass fraction from the central CO $1\rr0$
spectrum$^{\ref{elmfoot}}$.  Estimating the dynamical mass over the
central beam in a similar fashion for our data, i.e. assuming that the
maximum rotational velocity occurs at the edge of the
beam\footnote{Figure 9 of \citet{cjrb98} shows that this is a fair
  assumption in this particular case.}, gives $M_{\rm
  dyn}\approx3.1\times10^9$\Mo ~within the central $\la440$ pc,
compared to the $\approx3.9\times10^9$\Mo\ found by
\citet{ekhj98}. The molecular mass over the same region is $M_{\rm
  H_{2}}\approx4.4\times10^8$\Mo, giving, like \citet{ekhj98}, $M_{\rm
  H_{2}}/M_{\rm dyn}\approx 0.1$. The large increase in the gas mass
beyond this radius (previous paragraph), increases the gas mass
fraction by a factor of three (to $M_{\rm H_{2}}/M_{\rm dyn}\approx
0.3$) within 680 pc, thus being consistent with the presence of a
molecular ring in the central regions of Circinus. Note, however, that
an overestimate in the dynamical mass is possible when there are
non-circular motions present, due to the presence of non-axisymmetric
potential (e.g. a bar), particulalry in the central regions
\citep{kw02}. Therefore, the ring interpretation should only be
considered as one possibility. Finally, we see that
none of the central $1\rr0$, $2\rr1$ or $3\rr2$ spectra
\citep{cjrb98,cjhb99} exhibit the horns apparent in the global
spectra.
\begin{figure}
\vspace{6.4cm} 
\includegraphics{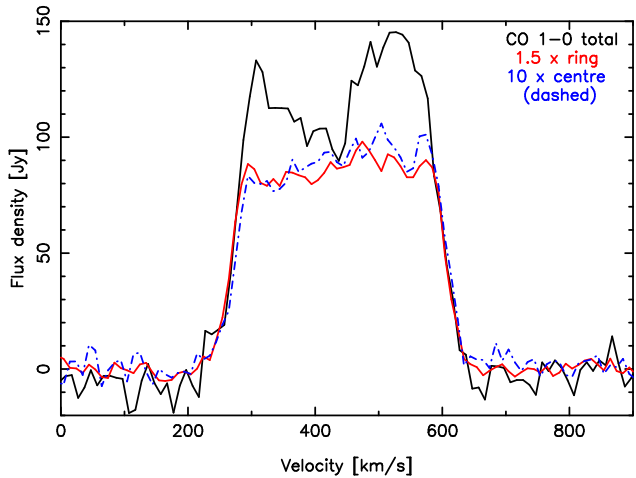}
\caption{Overlay of the global (black), ring region (red) and central
(dashed blue) CO $1\rr0$ spectra. All have been corrected by 27
Jy\,K$^{-1}$ and the ring region and central spectra scaled up by 
factors of 1.5 and 10, respectively.}
\label{horns}
\end{figure}
This difference is also seen in our data, where in Fig \ref{horns} we
overlay the global CO $1\rr0$ spectrum upon those of the centre position
(within the central $45\arcsec$ beam) and the molecular ring region
(within the central $\approx\pm40\times40 \arcsec$ mapped positions).
This shows that our large-scale CO maps detect the more extended
($\gapp1$ kpc) gas, which rotates at $\approx\pm130$ \kms\ (see
Sects. \ref{sect:trm} and \ref{rc}).

Converting the average CO $2\rr1$ velocity integrated line intensity
of $I_{\rm A} = 3.82 \pm 0.04$ K\kms, via $\eta_{\rm mb}=0.5\pm20\%$
at 230 GHz, gives $I_{\rm mb} = 7.7 \pm 1.5$ K\kms\ resulting in a
CO $2\rr1$ luminosity of $L_{{\rm CO}~2\rr1} = 350 \pm 70$
K\kms\,kpc$^2$. From previous observations of the central position
\citep{jabr91,cjhb99}, $\frac{{\rm CO}~2\rr1}{{\rm CO}~1\rr0}$
intensity ratios of close to unity were noted and Figs. \ref{cototal}
and \ref{horns} indicate that this ratio drops slightly in the
horns of the CO profile, which traces the extended gas beyond the ring.

\begin{figure*} 
\vspace{6.6cm}
\includegraphics{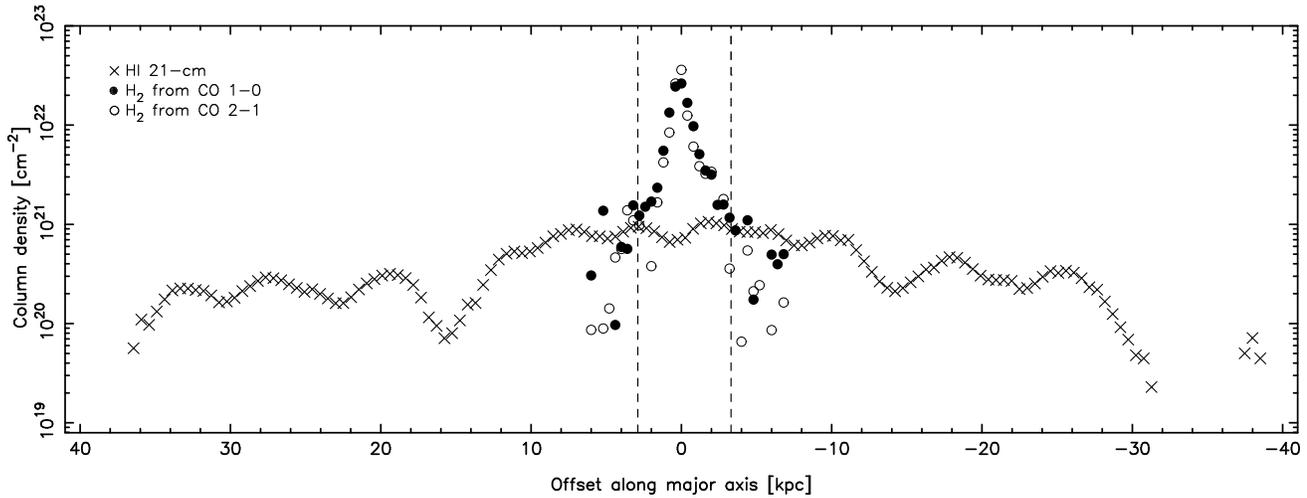}
\caption{\HI\ (crosses) and \MOLH\ (circles) column densities as a
  function of galactocentric radius (NE to SW along the major
  axis). The \HI\ column densities are calculated in the optically
  thin regime from a brightness temperature of 45.93 K\,Jy$^{-1}$ per
  the main-beam solid angle of $3.52 \times 10^{-7}$ sr. The
  \MOLH\ column densities have been calculated using $N_{{\rm
      H}_{2}}/I_{{\rm CO}} = 2.3 \times 10^{20}$ \scm/[K \kms]
  \citep{sbd+88}: For CO $2\rr1$ (unfilled circles), we use a
  main-beam efficiency of $\eta_{\rm mb} = 0.5$ and for CO $1\rr0$
  (filled circles), $\eta_{\rm mb} = 0.7$. The dashed lines represent
  the offsets of the last clear ($>3\sigma$) CO detections.}
\label{all}
\end{figure*}
The \MOLH\ (derived from CO $1\rr0$ \& $2\rr1$) and \HI\ column
densities along the major axis are shown in Fig.~\ref{all}. Since the
former is derived from both the CO $1\rr0$ and $2\rr1$ transitions
using the same $N_{\rm H_{2}}/I_{\rm CO}$ conversion ratio, it
represents a profile of the $\frac{{\rm CO}~2\rr1}{{\rm CO}~1\rr0}$
intensity ratio along the major axis.  Over the range where we have
clear detections (for \MOLH\ column densities of $N_{{\rm
    H}_{2}}\ga10^{21}$ \scm), we see that the high $\frac{{\rm
    CO}~2\rr1}{{\rm CO}~1\rr0}$ ratio holds. This may suggest that the
conditions ($T_{\rm kin} \sim 50 - 150$ K, $N_{\rm CO} \sim 0.2 - 20
\times 10^{18}$ and $n_{{\rm H}_{2}} \ga 10^3$ \ccm) found over the
central beam \citep{cjrb98,cjhb99,hak+08} could hold over the bulk of the
molecular gas, although these are high temperatures to maintain over a
3 kpc radius. Within this region we see very high molecular fractions of
$f\equiv2N_{\rm H_{2}}/[2N_{\rm H_{2}} + N_{\rm HI}]\approx1$, the
value of which drops to $f\la0.7$ beyond the limit of the clear CO
detection range ($\ga3$ kpc). That is, in the case of Circinus we
witness the limiting atomic gas column density of $N_{\rm
  HI}\la10^{21}$ \scm, which is typical for disk galaxies
(e.g. \citealt{ckbg94,rv96}), with gas above these column densities
being converted into the molecular state under the favourable
conditions close to the nucleus \citep{ys91,sch01}.

\subsection{Gas dynamics}
\label{dyn}

\subsubsection{Position angle variations}
\label{warps}

\begin{figure*} 
\vspace{5.6cm}
\includegraphics{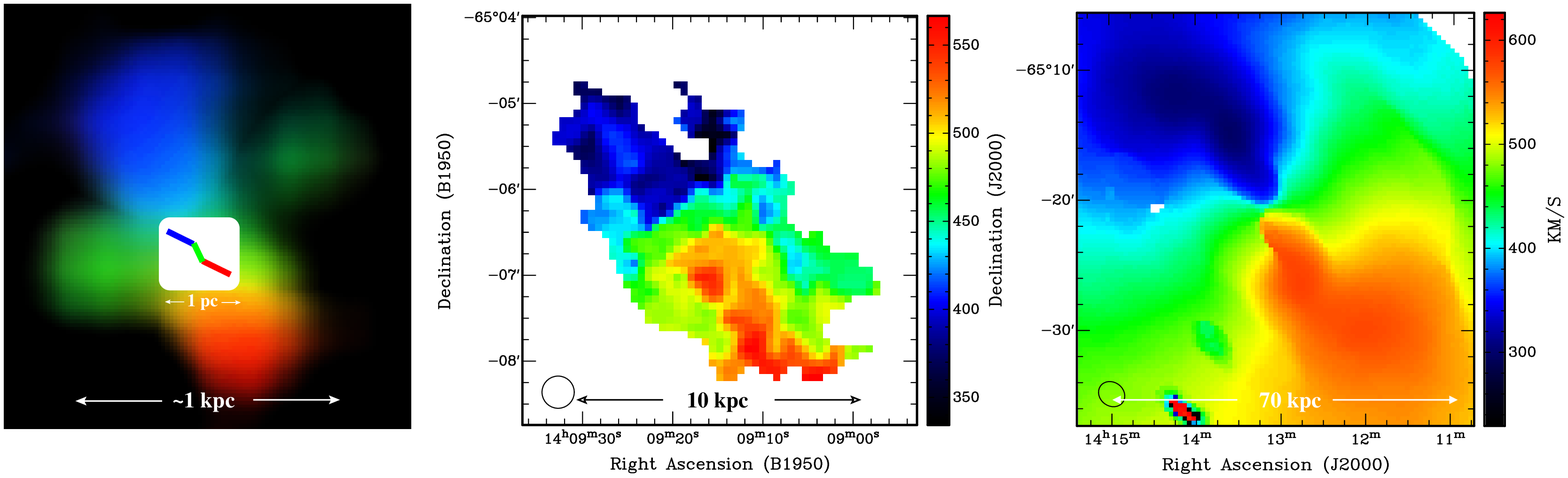}
\caption{The ``zig-zagging'' of the position angle on various scales
  in Circinus. Left: The molecular gas -- the inset shows the
  \WAT\ maser emission from the warped accretion disk in the inner
  parsec \citep{gbe+03}, which is superimposed upon the deconvolved
  (to a resolution of 10\arcsec\ using the method of \citealt{ryd08})
  map of the inner CO $2\rr1$ emission \citep{cur98}. This resolves
  the ring in the NE--SW diagonal, as well as the perpendicular
  outflow \citep{cjrb98,crjb98}. Middle: The CO $2\rr1$ first order
  moment map (due to binning of the data, the weaker extreme velocity
  channels are lost, see Sect. \ref{sect:trm}). Right: The central
  \HI\ first order moment map. In the latter two maps we show the beam
  in the bottom left corner, $22\arcsec$ and $124\arcsec \times
  107\arcsec$, respectively. In the colour version of the figure, blue
  is approaching, red is receding and green is systemic gas.}
\label{warp-scales}
\end{figure*}

Previously, \citet{jkeh98} noted two large spiral arms
originating at the end of the atomic gas bar at a radius of $\sim5$ kpc
and, as seen in the \HI\ distribution (\citealt{jkeh98} and
Fig.~\ref{large}), the bar itself is enclosed by a large ring of atomic gas.
Referring to the CO emission (Figs. \ref{co1} \& \ref{co2}), being
limited to $\lapp\pm3$\arcmin\ the molecular gas extends to only the
very central regions of the \HI\ (Fig.~\ref{large}, left). However,
similar to the \HI\ (Sect. \ref{himaps}), there appears to be an
extension in the more tenuous emission towards the SWW.
\citet{ekhj98} also noted this ``disturbance'' in the kpc-scale CO
emission, but since this was due to a single position spectrum, they
deemed it doubtful. We find, however, that feature is visible over
several spectra and at the peak has an integrated intensity of $I_{\rm
  A}= 20.3\pm1.3$ K \kms. This occurs at a distance of 1.8 kpc from
the centre position with a position angle of $PA\approx236$\degr, thus
exhibiting a ``warp'' from the $\approx214$\degr\ position angle of the
molecular ring (Fig.~\ref{warp-scales}). 
Such a change in position angle is also seen in the atomic gas (see also
\citealt{jkeh98} and Sect. \ref{sect:trm}) on the 10-kpc scale and
Fig.~\ref{warp-scales} illustrates that similar changes also occur at
the inner edge of the accretion disk. Although contained within the
central 0.1 pc, the accretion disk shares a similar position angle
($209\pm3$\degr) to the molecular and atomic rings, while the disk at
0.4 pc ($236\pm6$\degr) has the same position angle as the extended
molecular ($\ga1$ kpc) and atomic gases ($\ga10$ kpc).

These abrupt changes in position angle could be due to large
non-circular motions, perhaps caused by bar potentials in the inner
parts of the galaxy where the CO resides, as has been found for many
other spirals \citep{stcs93,rsv99,sois99}, including the Milky Way
\citep{fux99,dht01}. In these cases, the molecular gas is found to abruptly
change direction when transversing the dust lanes on the way to the
nuclear ring \citep{rsv99,dht01}, and in the case of Circinus, a dusty
nuclear bar \citep{mmoo94} nestling within the molecular ring (see
figure 1.7 of
\citealt{cur99})\footnote{http://nedwww.ipac.caltech.edu/level5/Curran/frames.html}
has been found. For Circinus it is apparent that a ring forms at
the ends of each bar and a tendency for strong bars to terminate in
rings has been noted in the dust structure of 75 galaxies by
\citet{pm06}.  Furthermore, as is seen for other galaxies
(e.g. \citealt{sois99}) and from numerical simulations of barred
potentials \citep{wad94,ba99,ab99,ks06}, the nuclear bar may provide a
means for transporting the gas to the nucleus of Circinus. Further
support is lent by the gas mass fraction of $M_{\rm H_{2}}/M_{\rm
  dyn}\approx 0.1$ within $\approx400$ pc, derived above, which is
consistent with the bar driven transport of molecular gas to the
nucleus \citep{sois99}. For Circinus, we therefore suggest, that upon
each collision with a bar or ring, the gas undergo a jolt which shifts the
position angle by $\approx20\degr$. As illustrated in
Fig.~\ref{warp-scales}, these same twists and turns in position angle
occur on wildly different scales and this may be the first case that
the orientations and kinks are seen to be repeated all of the way down
to the sub-pc accretion disk.

\subsubsection{Position--velocity diagrams}
\label{sect:pv}

\begin{figure*} 
\vspace{7.1cm}
\includegraphics{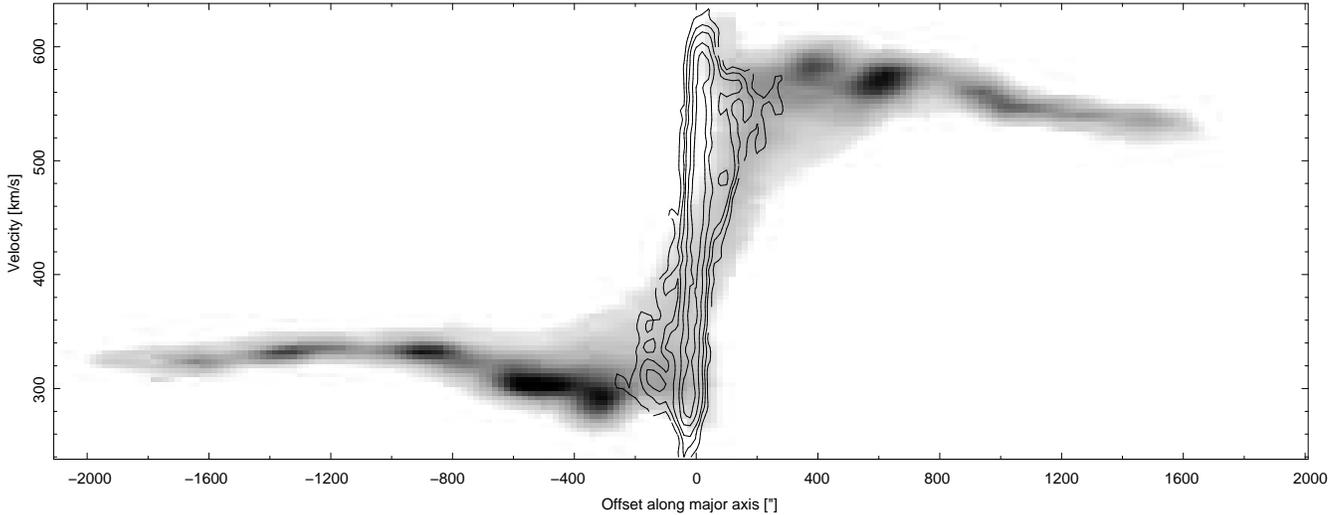}
\caption{\HI\ (grey-scale) and CO $1\rr0$ (contours) position-velocity diagrams of the 
Circinus galaxy along
  the major axis ($PA=210\degr$). The plot ranges from the NE to the
  SW.}
\label{pv}
\end{figure*}

In Fig.~\ref{pv} we show the \HI\ and CO $1\rr0$ position--velocity (p--v) diagrams of
Circinus along its major axis\footnote{Due to the lower resolution and
sensitivity to the inner regions, the diagram for the minor axis is
fairly featureless and the reader is referred to the p--v diagram of
\citet{jkeh98}.}. As expected, the \HI\ distribution is similar to
that of \citet{jkeh98}, with a rising component extending to a radius
of $\approx400$\arcsec\ ($\approx8$ kpc), being followed by a flatter component. We also
see, like \citet{jkeh98}, that along the major axis the gas is more
extended in the NE than the SW (also visible in Fig.~\ref{all}), being
clearly visible to 2000\arcsec\ ($\approx40$ kpc) and 1600\arcsec
($\approx30$ kpc) respectively.

\begin{figure} 
\vspace{6.4cm} 
\includegraphics{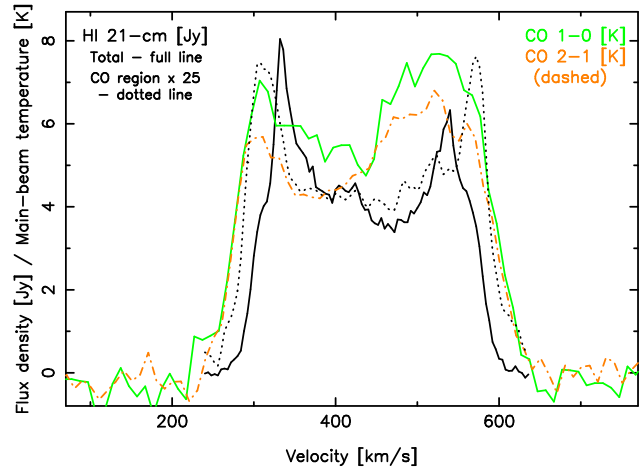}
\caption{Overlay of the global \HI\ (solid black), the \HI\ over the CO
  emission region (dotted black) and the CO ($1\rr0$ -- solid green \&
  $2\rr1$ -- dashed orange) emission profiles. The spectrum of the
  \HI\ over the CO emission region has been scaled up by a factor of
  25.}
\label{overlay}
\end{figure}

From the CO distribution it is apparent that the molecular gas traces
a different kinematical feature to the bulk of the atomic gas, in
which there is a significantly higher velocity gradient and a larger
velocity range, giving the difference in the global \HI\ and CO 
profile widths (Fig.~\ref{overlay}). 
High velocity wings in the CO distribution of other galaxies have
previously been attributed to high dispersion or rapid rotation of the
molecular gas in the inner disk \citep{sof92,ts99}. Zooming in on the
\HI\ emission profile over the CO emission region
(Fig.~\ref{overlay}), we see that this accounts for the weak large
velocity wings of the global \HI\ emission with a width close to that
of the CO profiles.  We therefore interpret this emission and the
wings of the CO profiles to the fastest part of the ring ($\gapp 130$
\kms), within the inner 600 pc \citep{cjrb98}.

The CO p--v distribution is very similar to that of
\citet{ekhj98}: Along the major axis we also observe `rigid' rotation
within the central $\pm$60\arcsec, as well as the flat rotation at
distances beyond this. The rigid rotation is consistent with the model
of the sub-kpc molecular ring, where the velocity increases to a
deprojected maximum of 180\kms\ at 400 pc \citep{cjrb98}. When
convolved with the 44\arcsec\ beam, this transforms to an observed
offset of $\approx$50\arcsec, thus suggesting that the rigidly
rotating inner $\pm$60\arcsec\ corresponds to the molecular
ring. Finally, the locations of the peak emission at velocities of
$294 \pm 2$\kms\ and $575 \pm 5$\kms, give symmetrical rotation about
$435 \pm 5$\kms, in excellent agreement with the systemic velocity
derived from the \HI\ (Sects. \ref{agm} and \ref{sect:trm}).

Beyond the inner $\pm$60\arcsec, the flat rotation along the major
axis is seen to occur at $\approx$300\kms\ (NE) and $\approx$560\kms\
(SW). Comparing these to the systemic velocity, gives a projected
velocity of about $\pm$130\kms, a value similar to that found by
\citet{ekhj98}. This corresponds to a deprojected velocity of
$\approx$130 to 150\kms\ (for $i$ = 78\degr\ to 60\degr,
Sect. \ref{comaps}), over a deconvolved distance of $\approx$800 pc to
4 kpc. The inner edge of this large-scale, solid body molecular disk
matches well with the inner rigidly rotating ring which has velocities
of $\la$150\kms\ at $\ga$600 pc \citep{cjrb98}.

\subsubsection{Tilted-ring model}
\label{sect:trm}

\begin{figure}
\vspace{14.3cm}
\includegraphics{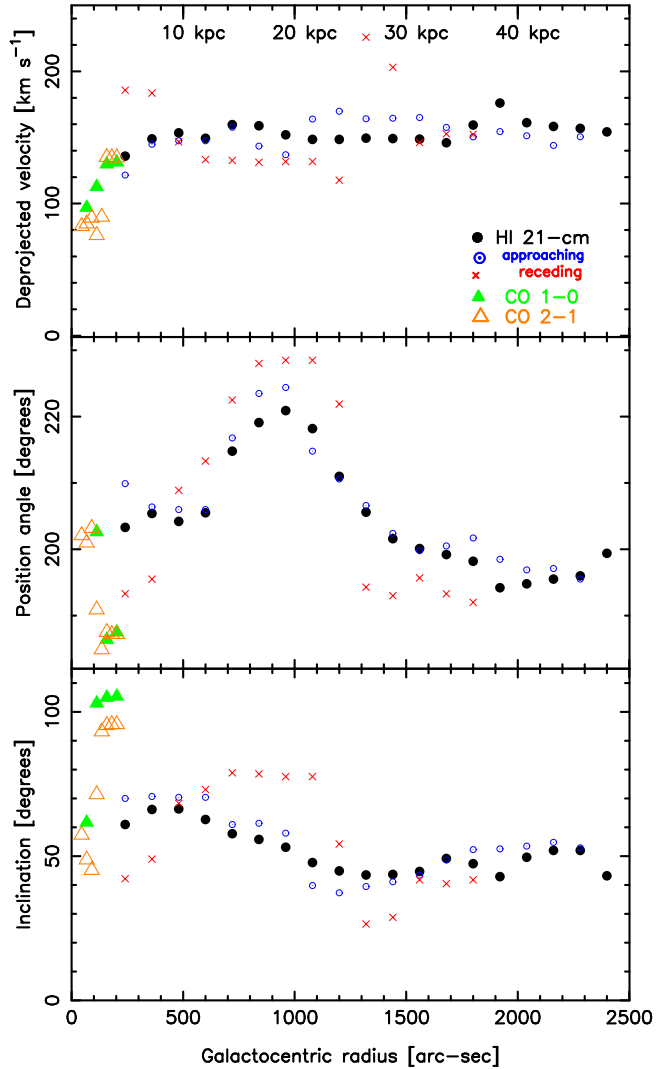}
\caption{The three parameter fit to the \HI\ (filled circles) and CO
  (triangles) velocity fields. The small circles show the model fit to
  the approaching side of the \HI, the crosses the receding side, the
  filled triangles the CO $1\rr0$ fit and the unfilled
  triangles the CO $2\rr1$ fit.}
\label{trm}
\end{figure}

In order to model the velocity field in Circinus a tilted ring model
was fitted to the first order moment maps. This was performed as per
the modelling of \citet{rlw74,beg89,tmpn97}, using with the {\sc aips}
task {\sf gal}, where by assuming that the velocity field was constant
over each 120\arcsec\ annulus ($\approx1\times$ the beam-size), we
fitted concentric annuli. In the first iteration, all six of the
fitting parameters were allowed to vary; the central position
coordinates, the position angle of the receding major axis, the
inclination of the orbital plane, the systemic velocity and the
maximum rotation velocity. The fits to the inner four annuli, from
120\arcsec\ to 600\arcsec, where the velocity field is most symmetric,
each gave the same central position and systemic velocity of
$\alpha_{\rm J2000} = 14^{\rm h}\,13^{\rm m}\,10^{\rm s}$,
$\delta_{\rm J2000}=-65\degr\,20\arcmin\,49\arcsec$ and $v_{\rm
  HI}=440$ \kms, respectively. That is, within 30\arcsec\ (a quarter
of the HPBW) and 1.5 \kms\ (half a channel) of the values obtained by
\citet{jkeh98}. Adopting the Jones et al. values to drive the tilted ring model,
we then ran another iteration of {\sf gal}, keeping these values fixed
and looping round the annuli in order to fit the full velocity
field. This was done out to 2460\arcsec, as beyond this the data
proved too poor to fit a model\footnote{Due to the blanking of some
  ``hot'' pixels in the receding side this is only modelled to
  $1800\arcsec$.}. This gave a three parameter fit very similar to
that of the initial model for the whole \HI\ velocity field, as well
as separate fits for the approaching and receding sides. The results
are shown in Fig.~\ref{trm}\footnote{The model velocity field and the residuals are available from http://www.phys.unsw.edu.au/$\sim$sjc/circinus/}.

To model the molecular gas, datacubes of the CO $1\rr0$ and $2\rr1$
were constructed in {\sc class}, in first order moment maps with
5-point hanning along the spectral axis. These were smoothed in order
to improve signal-to-noise ratio to a point where the data were
useful.  
We then
performed the modelling as for the \HI, fitting a rotation curve with
the constant velocity model and using the centre position and systemic
velocity derived from the \HI\ data. Again, we used annuli of
$1\times$HPBW width (45\arcsec\ for CO $1\rr0$ \& 22 \arcsec\ for
$2\rr1$), out to 200\arcsec\ and 170\arcsec, respectively, where the
data appeared reasonable, although none of the CO first moment maps
where of nearly as high a quality as for the \HI.

This is the first time that a tilted ring model has been applied to
the molecular gas in this galaxy and we see that for the inner
rotation curve (Fig.~\ref{trm}, top), the fit is consistent with the
\HI\ data, with the outer values of $\approx130$ \kms\ agreeing well
with the those derived by assuming circular orbits inclined at
$\gapp70$\degr\ (Sect. \ref{sect:pv}). Being restricted to radii of
$\gapp900$ pc, however, the model cannot trace the high velocities of
the inner molecular ring discussed above.

The CO model fits to the position angle are more awry, although
still within $\la20$\degr\ of the \HI\ model (Fig.~\ref{trm},
middle), but larger than the changes in position angle discussed in
Sect. \ref{warps}. For the inclination (Fig.~\ref{trm},
bottom), some CO fits are reasonably consistent
with the \HI\ model within the inner $160\arcsec$, but jump up to
$\approx100^{\circ}$ beyond this, at radii where the drop in position
angle is seen. We therefore attribute this to the poorer first moment
maps of the CO emission, which were more difficult than the \HI\ to
model, although, again, we do expect the inclination to rise within the inner
kpc (Table \ref{warp}).
\begin{table} 
\caption{Inclination angles, $i$, at various radii in Circinus. \label{warp}}
\begin{center}
\begin{tabular}{lccl} 
\hline 
Species & Radius    & $i$                       & Reference\\
\hline
\WAT & 0.1--0.4 pc  & edge-on                   & \citet{gbe+03} \\
CO   & 0.1--0.6 kpc & $78\degr \pm 1\degr$      & \citet{cjrb98} \\
     & 3 kpc        & $\ga63\degr \pm 5\degr$   & \citet{ekhj98}$^{\rm a}$\\
     & 3 kpc        & $60\degr \pm 6\degr$      & This paper$^{\rm b}$\\
H$\alpha$& 4 kpc    & $40\degr \pm 10\degr$      &\citet{ekj+98}\\
\HI  & 8 kpc      & $\approx70\degr$          & \citet{jkeh98} \\
     & 26 kpc       & $\approx40\degr$          & \citet{jkeh98} \\
     & 39 kpc       & $\approx60\degr$          & \citet{jkeh98} \\
     & 39 kpc       & $65\degr \pm 2\degr$      & \citet{fkl+77} \\
     & 47 kpc       & $\approx50\degr$          & This paper \\
\hline
\end{tabular}
\end{center}
{Notes: $^{\rm a}$\protect\citet{ekhj98} actually quote an inclination
angle of at least 73\degr, however we obtain 63\degr\ based on the CO
extent along the major and minor axes quoted. $^{\rm b}$By comparing
the scale sizes of the minor and major axes emission and applying,
like \protect\citet{ekhj98}, an uncertainty of one mapped position per
each full axis (see Sect. \ref{comaps}.)}
\end{table}

Comparing the \HI\ model with that of \citet{jkeh98}, we see that out
to a radius of 1700\arcsec, where their model was deemed reliable,
there is excellent agreement in all three parameters, with the
rotation curve, as well as all of the peaks and troughs in the
position and inclination angles, being reproduced.  Interestingly, the
offsets of the receding model fit from those of the approaching half
and the whole \HI\ velocity field are also reproduced over
$\approx1100-1800\arcsec$, confirming the asymmetry of the outer
Circinus disk as found by \citet{jkeh98}.  Beyond the inner
1700\arcsec (which is discussed in detail by \citealt{jkeh98}), both
the position and inclination angles settle close to the outer values
previously found. With regards to the rotation
curve, after one more small climb, this remains fairly flat,
indicating that the fastest rotating atomic gas is associated with the
large-scale diffuse emission and not within the sub-kpc ring
(Sect. \ref{sect:pv}).

\section{Masses in Circinus} 
\label{sect4}

\subsection{Dynamical masses}
\label{rc}

\begin{figure} 
\vspace{7.0cm} 
\includegraphics{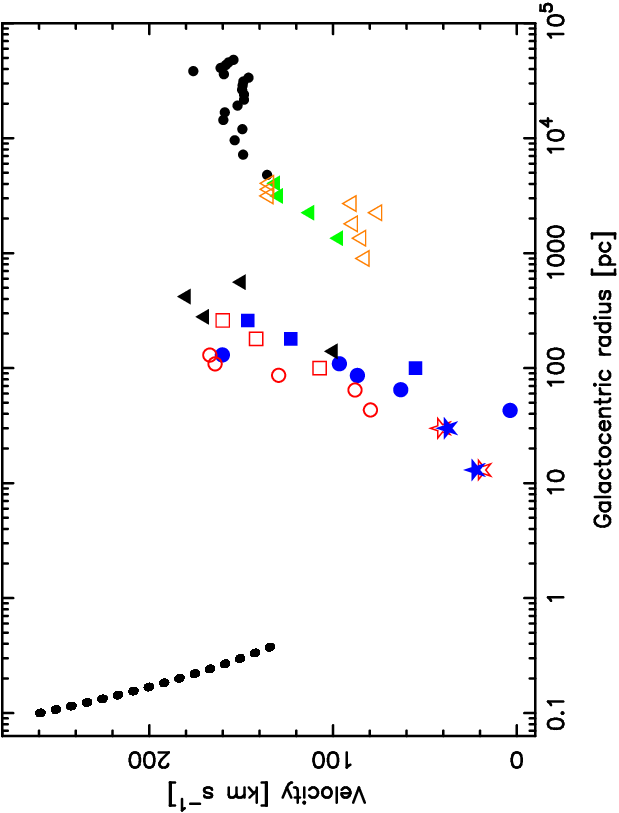}
\caption{The rotation curve over the major axis of the Circinus
galaxy. The dotted curve shows the Keplerian fit to the \WAT\ masers
\protect\citep{gbe+03}, the stars and squares show the stars and N{\sc
ii} emission line data \protect\citep{mktg98,osmm94}, respectively,
and the circles the show the \MOLH\ data of
\protect\citet{dfr+98}. The filled blue symbols represent the NE
(approaching) data and the unfilled red symbols the SW (receding) data
over these regions. Farther out, the symbols are as per
Fig.~\ref{trm}; the black triangles represent the CO ring model
\protect\citep{cjrb98}, the coloured triangles the larger-scale CO
($1\rr0$ -- filled green \& $2\rr1$ -- unfilled orange) and the small
circles the \HI.} \label{rot}
\end{figure}

In Fig.~\ref{rot} we combine the rotation curves from the tilted ring
models with those from the literature to show the dynamics
over the whole galaxy.
The overall profile is fairly typical for spirals (type Sa -- Sc) with
a steep initial rise within the inner few kpc, beyond which the
velocity plateaus
\citep{rft80,rftb82,rbft85,bos81,bos81a}. 

\begin{figure}
\vspace{7.0cm}
\includegraphics{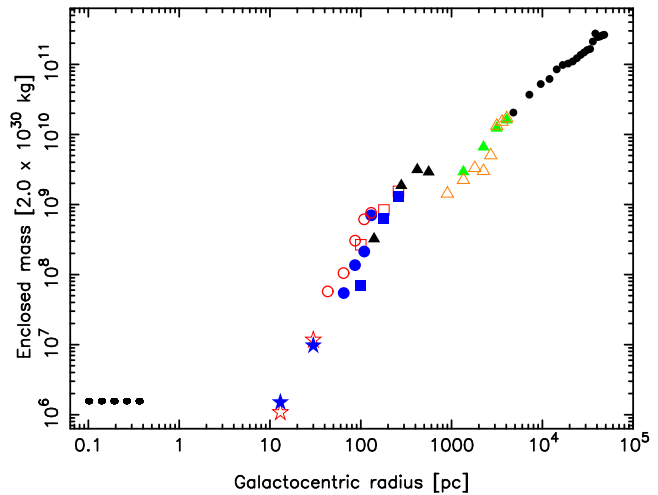}
\caption{The dynamical mass distribution of the Circinus galaxy. The 
   symbols are as per Fig.~\ref{rot}. This confirms the underestimate in the
velocities from the CO $2\rr1$ tilted ring model, although the outer three
$1\rr0$ estimates appear quite reasonable.}
\label{mass}
\end{figure}

Specifically, in the inner 0.4 pc we see Keplerian rotation indicative
of dynamics which are dominated by a compact massive object (see
\citealt{gbe+03}). Beyond this, from $\approx10$~pc to
$\approx400$~kpc the velocity scales directly with
distance\footnote{There is some scatter, particularly at the lower
radii, due to the effects of seeing and in Figs.  \ref{rot} and
\ref{mass} we have therefore excluded the central three points of
\protect\citet{dfr+98} [where $r \la 1\arcsec$, cf. a seeing of
$\ga$1.4\arcsec] and the central point of \protect\citet{mktg98}
[where the seeing is $\approx$0.5\arcsec\ -- 0.6\arcsec]. As well as
the seeing, uncertainties in the velocities are quoted: 9 to 15\kms\
for the stars and 30\kms, for the N\,{\sc ii} emission \citep{mktg98},
which we do not show for the sake of clarity. While these data are
characterised by linear fits on linear plots, in the log plot
uncertainties in the position are hugely exaggerated.}, despite both
the \MOLH\ and CO intensity dropping significantly over the inner few
hundred pc (\citealt{dfr+98} and \citealt{cjrb98}, respectively),
probably due to consumption by the  starburst activity within the central kpc
\citep{mmoo94,ekj+98}.

The outer edge of the starburst/molecular ring is marked by a sharp
drop in the rotational speeds, although probably not to the extent as
indicated by the CO $2\rr1$ tilted ring model\footnote{Also, the
decrease would be much more gentle with radius than evident from the
logarithmic scale of the plot.}, which is confirmed by the dynamical mass distribution 
(Fig.~\ref{mass}). 
Beyond the molecular ring, velocities settle at $\approx150$ \kms,
where the gas follows the general, large-scale solid body rotation
typical of disk galaxies, although Fig.~\ref{mass} may suggest that
there is not much additional (gravitating) matter between the ring and
the large-scale atomic gas. Beyond this plateau, the mass of the
galaxy clearly keeps climbing beyond the outer edge of the disk,
although with less of a gradient than the inner kpc, confirming the
differences in the p--v distributions between the atomic and molecular
gases (Fig.~\ref{pv}).  Visually, there is a drop of 3 magnitudes
within the central 1~kpc and with a Holmberg radius of 10 kpc
\citep{fkl+77} [see also Sect. \ref{himaps}], it is clear that the
halo must be dominating the outer dynamics of Circinus.

\subsection{Gas mass fractions}

In Table \ref{masses} we summarise the various gas and dynamical masses
found for the Circinus galaxy.
\begin{table} 
\caption{Gas and dynamical masses at various radii in Circinus. \label{masses}}
\begin{center}
\begin{tabular}{crcl} 
\hline
         Mass [\Mo]      &   Radius   & Type            & Reference \\
\hline
\hline
$\approx4\times10^8$   & $<440$ pc  & \MOLH$^{\rm a}$ & \citet{ekhj98}\\
$\la4.7\times10^9$       & $<440$ pc  & Total$^{\rm b}$ &  \citet{ekhj98}\\
\hline
$1.0\pm0.2\times10^9$    & $<680$ pc  & \MOLH$^{\rm a}$ & This paper \\
$3.3\pm0.3\times10^9$    & $<560$ pc  & Total           & \citet{cjrb98}\\
\hline
$2.0\pm0.5\times10^9$    & $\la5$ kpc & \MOLH$^{\rm a}$ & This paper \\
$\approx2.1\times10^{10}$& $\la5$ kpc & Total           & This paper\\
\hline
$1.9\pm0.1\times10^9$    & $<10$ kpc  & \HI             & This paper\\
$\approx6.2\times10^{10}$ & $<12$ kpc  &  Total          & This paper\\
\hline
$\geq3.8\times10^9$      & $<35$ kpc  & \HI             & \citet{jkeh98}\\
$1.9\pm0.3\times10^{11}$ & $<35$ kpc  & Total           & \citet{jkeh98}\\
\hline
$7.2\pm0.5\times10^9$    & $<40$ kpc  & \HI             & \citet{fkl+77}\\
$1.3\pm0.2\times10^{11}$ & $<40$ kpc  & Total           & \citet{fkl+77}\\
\hline
$\approx8.1\times 10^9$  & $<60$ kpc  & \HI             & \citet{jsk+00}\\
$2.6\pm0.6\times 10^{11}$  & $<50$ kpc  & Total$^{\rm c}$ & This paper\\
\hline
\end{tabular}
\end{center}
{Notes: $^{\rm a}$Applying the $N_{{\rm H}_{2}}/I_{{\rm CO}}$
 conversion ratio of $2.3\times10^{20}$ \scm/[K \kms] \citep{sbd+88}
 [see also Sect. \ref{gas}]. $^{\rm b}$CO using $i = 63\degr$ (Table~\ref{warp}) [\citet{ekhj98} quote $\la3.9\pm0.2\times10^9$\Mo ~for $i
 = 73\degr$]. $^{\rm c}$assuming that the uncertainty in the tilted
 ring model, which is not given by {\sf gal}, is much less than the
 uncertainty due to the distance estimate of Circinus (see footnote
 \ref{foot}).}
\end{table}
 \citet{cjrb98} reported a gas mass fraction of $M_{\rm H_{2}}/M_{\rm
dyn}\approx0.5$ within the central 560 pc, which called into question
the Galactic $N_{\rm H_{2}}/I_{\rm CO}$ conversion ratio of
$\approx2.3 \times 10^{20}$ \scm/[K \kms] \citep{sbd+88}. We see,
however, that this does not apply over the whole of the galactic disk,
falling to the nominal $\la0.1$ \citep{mb81} beyond the molecular
ring, and upon reaching the Holmberg radius, the gas mass fraction
falls to $\approx0.03$. Combining the atomic gas mass
(\citealt{jsk+00}) with the molecular gas mass gives $M_{\rm
gas}\equiv M_{\rm HI} + M_{\rm H_{2}}\approx1\times10^{10}$\Mo, and so
$M_{\rm gas}/M_{\rm dyn}\approx0.03$ holds over the inner $\approx50$
kpc of the galaxy (Table \ref{masses}). This is significantly smaller
than the value obtained by \citet{ekhj98} [$M_{\rm gas}/M_{\rm
dyn}\approx0.6$], and is consistent with the SAS3 (Hubble type Sb,
where $M_{\rm gas}/M_{\rm dyn}\sim0.08$, \citealt{ys91})
classification of Circinus \citep{ddc+91}.

\subsection{Mass inventory}

In Table \ref{inv} we give the global values for the various masses
in Circinus.
\begin{table} 
\caption{Mass inventory of the Circinus galaxy. The masses quoted are
         the (most) global values.\label{inv}}
\begin{center}
\begin{tabular}{lcr} 
\hline
                            & Mass [\Mo]                & Radius \\
\hline
Molecular gas        & $2.0\pm0.5\times10^{9}$   & $\la5$  kpc \\
Atomic gas            & $\approx8.1\times 10^9$  & $\la60$ kpc \\
Ionised gas    &  $\sim10^6$              & $\la0.3$ kpc \\
Stars                       &  $\sim10^{11}$       & $\la10$ kpc\\
Dust                   &  $\sim10^6$                 & $\la60$ kpc\\
\hline
Total                       & $\sim10^{11}$ & $\la60$ kpc \\
\hline
\hline
Dynamical mass              & $\approx3\times10^{11}$ & $<50$ kpc\\
\hline
\end{tabular}
\end{center}
\end{table}
The steep decline in column density beyond the central 3 kpc
(Fig.~\ref{all}), suggests that we have mapped most of the molecular
gas, although from the gas-to-dust ratio, \citet{sb93} estimate
$M_{\rm H_2} \approx0.6 \times M_{\rm HI}$, which would give $M_{\rm
H_2} \approx5 \times 10^9$\Mo\ using the total atomic gas mass of
\citet{jsk+00}. The ionised gas mass is obtained from the integrated
H$\alpha$ flux of the ionisation cone within the central $30\times30$
arcsec$^2$ \citep{ekj+98}. Since most of the ionised gas appears to
reside in the ionisation cone and star-burst ring, within a radius of
$\approx300$ pc \citep{mmoo94}, we do not envision the total ionised
gas mass competing with that of the neutral gas (recently,
\citealt{rga08} estimate an ionised gas mass of between $3\times10^3$
\Mo\ and $1\times10^6$ \Mo\ within a radius of 0.7 kpc). For the
stellar mass content we have simply used the value typical for a large
spiral and the dust mass is estimated from the gas-to-dust ratio given
by \citet{sb93}. From this, it is apparent that the stellar population
could easily account for the observed dynamical mass, although, as
discussed in Sect.~\ref{rc}, the blue and visual brightnesses have all
but petered out by time a radius of 10 kpc is reached.  At this radius
the dynamical mass is $M_{\rm dyn}\approx5\times 10^{10}$\Mo, with
$\approx10^{11}$\Mo\ being reached within the central 20 kpc
(Fig.~\ref{mass}).

\section{Summary} 

We have studied the atomic and molecular emission in the Circinus
galaxy through large-scale maps of the \HI\ and CO $1\rr0$ \& $2\rr1$
transitions.  Our \HI\ mosaic represents the largest high resolution
image of this galaxy to date, in which we recover 70\% of the total
\HI\ flux, while resolving the detailed gas distribution. We have
mosaiced a field of $\approx\pm1\degr\times\pm1\degr$, thus recovering
$\approx30\%$ more flux than the primary beam limited image of
\citet{jkeh98}, giving the atomic gas distribution and dynamics to a
radius of 50~kpc.  The CO maps also represent the largest to date, spanning
$\pm260\arcsec$ and $\pm100\arcsec$ along the major and minor axes,
respectively, at a sensitivity of $\approx10$ mK per each 10 \kms\ channel.
From our observations:
\begin{enumerate}

\item We confirm that the atomic gas in Circinus is distributed in a
$\ga50$ kpc irregular disk, which is a factor of $\ga5$ times as
extended as the optical and infrared emission. A tilted ring model of
the \HI\ velocity field agrees very well with the known warping in
position angle and inclination, found by \citet{jkeh98}, out to
$\approx30$ kpc. Beyond this (to 50 kpc), the warping appears to
dampen significantly.

\item We obtain clear detections of CO at a galactocentric radius of
3~kpc. Although the position angles and inclinations yielded by the
tilted ring model of the CO velocity field may be unreliable, by
assuming circular orbits for the molecular gas we find that the
molecular disk also undergoes a ($\approx20\degr$) warp within the
inner 3~kpc, a radius at which it shares a similar inclination to the
atomic gas. Note, however, that the apparent warp may be the result
of non-circular motions, which are a distinct possibility due to
the presence of the atomic bar on these scales \citep{jkeh98}.

\item The inner part of each gas structure also appears to
maintain the same position angle on all scales; sub-pc (\WAT), sub-kpc
(CO) and $\lapp10$~kpc (\HI). This may suggest a continuous
transportation of material to the active galactic nucleus, while
``kinks'' are introduced when the outer diffuse gas encounters a bar.

\item We confirm the finding of the ring model of \citet{cjrb98},
that there is more molecular (and less atomic) gas in the receding (SW) half
of the ring.

\item We find that the CO profile is wider than that of the \HI\ in both
the full-width half maximum and the total velocity spread. A
comparison of the \HI\ and CO position-velocity diagrams confirms that
the molecular gas has different kinematics to the atomic gas.  We
believe that this indicates that the fastest rotating gas, which
occurs within the sub-kpc ring, is indeed primarily
molecular.

\item We also investigate the global integrated atomic and molecular
gas intensities, and find:

\begin{enumerate}
  \item Circinus has a total molecular gas mass of $M_{\rm H_{2}} \approx 2\times 10^{9}$\Mo, cf.
    the atomic gas mass of $M_{\rm HI}\approx8 \times 10^9$\Mo\ \citep{jsk+00}.
\item Previously, \citet{cjrb98} found a gas mass fraction of
     $M_{\rm gas}\approx M_{\rm H_{2}}\approx 0.5 M_{\rm dyn}$, within
     the central kpc, which cast into doubt the validity of applying the
     standard Galactic conversion ratio of $N_{\rm H_{2}}/I_{\rm CO}
     \approx 2 \times 10^{20}$ \scm/[K \kms] \citep{sbd+88}. However,
     over the global molecular gas emission we obtain a more canonical
     ($M_{\rm gas}\la0.1 M_{\rm dyn}$, \citealt{mb81}) ratio of
     $\approx0.03$. This is much smaller than the value ($M_{\rm gas}/M_{\rm
     dyn}\approx0.6$) obtained by \citet{ekhj98}. 

\item Within the inner $\approx$3 kpc, the gas remains predominantly
     molecular, with the molecular hydrogen column density ranging
     from $N_{\rm H_{2}} \approx 2\times 10^{21}$ to $\approx 4 \times
     10^{22}$ \scm\ cf. the atomic hydrogen column density of $N_{\rm
     HI} \la 10^{21}$ \scm.  Furthermore, the CO $2\rr1/1\rr0$
     intensity ratio of $\approx$1 previously observed over the
     central beam is also seen over the whole of this region.
  
  \end{enumerate}

 \item From the rotational dynamics of the gas, the steepest gradient
   of enclosed mass, $dM/dr$, occurs from $\approx$10 -- 400~pc,
   although there is a low abundance of gas in this region (again,
   $M_{\rm gas}\la0.1 M_{\rm dyn}$), probably due to  consumption by the
   starburst. At $\ga400$~pc, which is approximately coincident with
   the molecular ring ($M_{\rm gas}/M_{\rm dyn}\approx0.5$), there is
   a steep drop in velocity. This continues to $\approx1$~kpc, the
   extension of the atomic ring (\citealt{jkeh98}). Beyond these
   radii, velocities climb once more, until settling into solid body
   rotation at $\approx150$ \kms. This continues to the 50~kpc limit
   of our tilted ring model, although most of the blue, visible and
   infrared radiation emanates from the central 10~kpc, suggesting
   that the dynamics beyond this region are halo dominated.

\end{enumerate}

\section*{Acknowledgments}
We would like to dedicate this paper to the memory of Lars E B
Johansson (1945--2008), who worked extensively on Circinus supervising
SJC's doctorate, as well as providing valuable input to this
article. He will be sorely missed as a colleague and friend.  We
acknowledge the contribution of Raymond Haynes who was involved in the
early stages of this project as well as the ATCA \HI\ observations.
We wish to thank the anonymous referee for their prompt and
helpful comments, the SEST operators Francisco ``Pancho Pollo'' Azagra
and Felipe ``El Maestro'' Mac-Auliffe, as well as Rob Beswick, Roy
Booth, Bi\"{o}rn Nilsson, Per Bergman, Alan Pedlar, Carole Mundell,
Steve Longmore, Michael Burton, Martin Thompson and Roberto Maiolino
for their helpful input.  Thanks to Gustaf Rydbeck who originally
produced the deconvolved CO $2\rr1$ image (Fig.~\ref{warp-scales},
left).  Very special thanks to Cormac ``Mopra Boy'' Purcell for
converting the {\sc xs} FITS files into datacubes, which could then be
modelled. This research has made extensive use of the {\sc miriad},
{\sc xs} and {\sc karma} packages, the NASA/IPAC Extragalactic
Database (NED) which is operated by the Jet Propulsion Laboratory,
California Institute of Technology, under contract with the National
Aeronautics and Space Administration and NASA's Astrophysics Data
System Bibliographic Services.

\bsp

\label{lastpage}

\end{document}